\documentclass[fleqn,usenatbib]{mnras}
\usepackage{mathptmx}
\usepackage[varvw]{newtxmath}  
\usepackage[T1]{fontenc}
\usepackage{ae,aecompl}
\usepackage{times}
\usepackage{amsmath}
\usepackage{upgreek}
\usepackage{xcolor}
\usepackage{graphicx}
\usepackage{pdflscape}
\usepackage{multicol} 
\usepackage{amsmath,bm}
\usepackage{caption}

\begin{document}

\title{The post-gas expulsion coalescence of embedded clusters as an origin of open clusters}

\author[J. W. Zhou]{
J. W. Zhou \thanks{E-mail: jwzhou@mpifr-bonn.mpg.de}$^{1}$
Sami Dib \thanks{E-mail: dib@mpia.de}$^{2}$
Pavel Kroupa \thanks{E-mail: pkroupa@uni-bonn.de}$^{3,4}$
\\
$^{1}$Max-Planck-Institut f\"{u}r Radioastronomie, Auf dem H\"{u}gel 69, 53121 Bonn, Germany\\
$^{2}$Max Planck Institute f\"{u}r Astronomie, K\"{o}nigstuhl 17, 69117 Heidelberg, Germany\\
$^{3}$
Helmholtz-Institut f{\"u}r Strahlen- und Kernphysik (HISKP), Universität Bonn, Nussallee 14–16, 53115 Bonn, Germany \\
$^{4}$
Charles University in Prague, Faculty of Mathematics and Physics, Astronomical Institute, V Hole{\v s}ovi{\v c}k{\'a}ch 2, CZ-180 00 Praha 8, Czech Republic 
}

\date{Accepted XXX. Received YYY; in original form ZZZ}

\pubyear{2024}
\maketitle

\begin{abstract}
The mismatch between the mass function of the Milky Way's embedded clusters (ECs) and that of open clusters (OCs) raises the question of whether each OC originates from a single EC. In this work, we explore a scenario in which OCs form as a result of post-gas expulsion coalescence of ECs within the same parental molecular cloud. We model this process using N-body simulations of ECs undergoing expansion due to gas expulsion. Our initial conditions are based on the observed spatial, kinematic, and mass distributions of ECs in three representative massive star-forming regions (MSFRs). Initially, ECs are isolated. 
After further expansion, interactions between ECs begin, mutually influencing their evolution. We examine this process as a function of gas expulsion timescales, spatial separations between ECs, and their relative velocities. Our results demonstrate that, within a reasonable range of these parameters, the coalescence of ECs is robust and largely insensitive to initial conditions.
The mass of ECs plays a critical role in the coalescence process. More massive ECs form stable gravitational cores, which greatly facilitate coalescence and help the resulting cluster resist expansion and Galactic tidal forces. Additionally, the number of ECs also enhances coalescence. The current mass distribution of clumps in the Milky Way suggests that directly forming massive ECs is challenging. However, the coalescence of multiple low-mass ECs can account for the observed parameter space of OCs in the Milky Way.
\end{abstract}



\maketitle 

\section{Introduction}

Massive star-forming regions (MSFRs) are a major contributor to star formation in the Galaxy. The young stellar clusters that form within these regions may eventually develop into open clusters \citep{Lada2003-41, Motte2018-56}. It has been suggested that most, if not all, observed stars originate from embedded clusters \citep{Kroupa1995a-277, Kroupa1995b-277, Lada2003-41, Kroupa2005-576, Megeath2016-151, Dinnbier2022-660}. Despite their importance, the properties of embedded clusters and their relationship to open clusters remain unclear. The mechanisms driving the fragmentation of gas that forms these clusters, as well as the subsequent dynamics of their stars, are not yet well understood \citep{Megeath2016-151, Dib2023-524, Li2024MNRAS-532, Zhou2024-686}.
Although embedded clusters are the likely precursors to open clusters, \citet{Lada2003-41} discovered that only 7\% of embedded clusters survive the gas dispersal phase. In N-body simulations of \citet{Kroupa2001-322} and \citet{Brinkmann2017-600}, massive clusters containing many O stars most likely survive as open clusters. However, low-mass embedded clusters or groups dissolve quickly, partly owing to the loss of their residual gas.

In \citet{Zhou2024sub}, we collected samples of Galactic clumps, embedded clusters and open clusters to compare their physical properties. We showed that the mass distribution of open clusters covers a significantly larger mass range than that of embedded clusters, by about one order of magnitude. Given the current mass distribution of clumps in the Milky Way, the evolutionary sequence from a single clump evolving into an embedded cluster and subsequently into an open cluster cannot account for the observed open clusters with old ages and high masses. This fact is also supported by N-body simulations of individual embedded clusters. In order to explain the mass and radius distributions of the observed open clusters, embedded clusters with initial masses of more than 3000 M$_{\odot}$ are necessary. However, for the embedded cluster sample collected in \citet{Zhou2024sub}, the upper limit of the mass is less than 1000 M$_{\odot}$. Additionally, only a few clumps have masses larger than 3000 M$_{\odot}$. For an open cluster, its progenitor should have a significantly larger mass. Thus, the currently observed clumps cannot be the "direct" precursors of the currently observed open clusters with relatively older ages and larger masses. 

A possibility is that those massive and relatively old open clusters form from the post-gas expulsion coalescence of multiple embedded clusters. In \citet{Zhou2024sub}, we compared the separation of open clusters and the typical size of molecular clouds in the Milky Way, and found that most molecular clouds may only form one open cluster. Since a molecular cloud typically includes many embedded clusters, this result supports the coalescence scenario. Moreover, the typical separation between embedded clusters in MSFRs can be $\approx$1 pc \citep{Zhou2024-688}. Therefore, after these clusters expand, they should be able to undergo coalescence.
There is now extensive literature arguing that star clusters form through the merging of subclusters, both from simulations \citep{Vazquez2009-707, Fujii2012-753, Vazquez2017-467, Howard2018-2,Sills2018-477,Fujii2022-514,Dobbs2022-517,Guszejnov2022-515,Cournoyer2023-521,Reina2024arXiv,Polak2024-690} and observations \citep{Sabbi2012-754,Dalessandro2021-909,Pang2022-931,Della2023-674,Fahrion2024-681}. 
This study will examine the physical processes involved in the coalescence of embedded clusters after gas expulsion and expansion, and utilize simulations to fit the observed physical parameters of open clusters. This paper is organized as follows: in \S.~\ref{sampleclusters}, we summarize the sample of OCs that is used in this work. In \S.~\ref{res}, we present our results. In the first subsection (i.e., \S.~\ref{mergerstages}), we summarize the different phases at which the the mergers of OCs progenitors could possibly occur and in the subsection that follows, we present and discuss the results from numerical simulations that we employ in order to simulate this process. In \S.~\ref{discuss}, we discuss our findings and in \S.~\ref{conc}, we present our conclusions.

\section{Sample}\label{sampleclusters}

\citet{Hunt2023-673} performed a blind, all-sky search for open clusters, utilizing 729 million Gaia DR3 sources down to a magnitude of G$\approx$20. This effort resulted in a homogeneous catalog comprising 7167 clusters. By assessing the masses and Jacobi radii of these clusters, the catalog's members were classified into two categories: bound open clusters (OCs) and unbound moving groups (MGs) \citep{Hunt2024-686}. There are 5647 OCs and 1309 MGs, with 3530 OCs and 539 MGs classified as high-quality, based on median color-magnitude diagram (CMD) classifications of greater than 0.5 and signal-to-noise ratio (S/N) of greater than 5$\sigma$ \citep{Hunt2023-673}.
In this work, we only focus on the high-quality sample of OCs.
To calculate the photometric masses of the clusters, \citet{Hunt2024-686} first derived the photometric masses of the member stars in each cluster using the PARSEC isochrone \citep{Bressan2012-427}. They then corrected for selection effects and applied a correction for unresolved binary stars. Subsequently, mass functions were fitted and integrated to determine the total cluster mass.

\section{Results}\label{res}

\subsection{Mergers at different stages}\label{mergerstages}

During different stages of cluster formation and evolution, various types of mergers can occur.

\subsubsection{Formation of clumps}
This stage should be dominated by the hydrodynamics of the gas. Clumps are locally dense gas structures in molecular cloud, which serve as local star-forming sites within molecular cloud. 

\subsubsection{Formation of embedded clusters}
Once clumps form, their evolution and internal star formation proceed together until the formation of embedded clusters. 
There may be an isolated evolution from clumps to embedded clusters. Clumps are dynamically decoupled from their parent
molecular clouds \citep{Zhou2023-676, Peretto2023-525}.
As discussed in \citet{Urquhart2022-510},
the differences observed for the physical parameters (such as radius, mass, density) of clumps at different evolutionary stages are in fact the result of observational biases, rather than the evolution of the embedded objects. 
This implies that the clumps remain stable throughout their evolutionary process and do not merge. It also suggests that the material lost through outflow and accretion onto stars is roughly balanced by the inflow of new material onto the clumps during star formation.
\citet{Zhou2024-688} used the minimum spanning tree method to measure the separations between embedded clusters or very young star clusters and clumps in MSFRs investigated in the Massive Young Star-Forming Complex Study in Infrared and X-ray (MYStIX) project \citep{Feigelson2013-209, Kuhn2015-802}. 
The separations between clusters, between clumps, and between clusters and clumps are comparable, which indicates that the evolution from clump to embedded cluster proceeds in isolation and locally, and does not affect the surrounding objects significantly. These findings align with the conclusions of \citet{Zhou2024-682-173,Zhou2024-682-128}, which indicate that feedback from embedded clusters does not notably alter the physical characteristics of the surrounding dense gas structures. Using Gaia DR2 data, \citet{Kuhn2019-870} investigated the kinematics of subclusters identified in \citet{Kuhn2014-787} for the same MSFRs, and found no evidence that these groups are merging.  

\subsubsection{Evolution of embedded clusters}
As the effects of feedback from the embedded clusters becomes increasingly stronger, they are responsible for gradually dispersing the gas of the clumps \citep{Dale2008-391,Dale2012-424,Dib2011-415,Dib2013-436,Zhou2022-514,Lewis2023-944}. Due to the gas expulsion, the embedded clusters will expand and subsequently interact and coalescence. 
Recent observational progress has highlighted the crucial role of early expansion in the evolution of young star clusters. By leveraging Gaia DR2 and DR3 data in combination with multi-object spectroscopy, studies have revealed expansion processes occurring in very young clusters within star-forming regions
through both case studies \citep{Wright2019-486, Cantat2019-626, Kuhn2020-899, Lim2020-899, Swiggum2021-917, Lim2022-163, Muzic2022-668, Das2023-948, Flaccomio2023-670} and statistical works \citep{Kuhn2019-870, Della2024-683, Wright2024-533,Jadhav2024-687}.

Seven of the 17 MSFRs investigated in the MYStIX project are covered in the ATLASGAL survey \citep{Schuller2009-504}, which are the Lagoon nebula, NGC~6334, NGC~6357, the Eagle nebula, M~17, the Carina nebula, and the Trifid nebula.
As shown in Appendix.~A of \citet{Zhou2024sub}, the ATLASGAL+Planck 870 $\mu$m data \citep{Csengeri2016-585} was used to trace the gas distribution around embedded or very young star clusters. 
On the maps, clusters and clumps are well separated. In stage3, feedback from the newly formed embedded clusters will disperse the surrounding gas, then these clusters will be located within the cavities of HII regions. Finally, the clusters or their coalescence remnants will be gas-free. In this work, we only focus on the coalescence of the formed embedded clusters following their expansion, which occurs in stage3.

The dynamics of both stars and gas have an important influence on the structure and evolution of embedded star clusters. A comprehensive simulation should combine both components. However, considering the limitations of computational resources, we have to simplify the simulation by focusing only one the first aspect, namely that of stellar dynamics. 
The complexity of the simulation arises from the interaction between the cluster and the surrounding gas during the formation and evolution of the cluster. If the evolution of the cluster is relatively independent of the gas, as in stage3, we can primarily focus on N-body simulations. 
\citet{Sills2018-477} presents simulations of embedded clusters that model the simultaneous evolution of both the stellar and gas components, beginning with initial conditions that are motivated and directly drawn from the MYStIX observations in \citet{Kuhn2014-787}.
By using live gas particles, they account for the gravitational interactions between the gas and stars within these clusters. Their results suggest that the subsequent evolution of these regions is largely insensitive to variations in the parameters within reasonable ranges. The initially highly sub-structured systems rapidly evolves into spherical, monolithic, and smooth configurations, both in terms of spatial and velocity distributions. The evolution of young embedded clusters appears to be primarily driven by gravitational interactions among the stars, with the surrounding gas having only a minor influence.

\subsection{Coalescence of embedded clusters in simulations}\label{merge}

\begin{table*}
\centering
\caption{"Terminated time" is the time when the simulation is terminated. The cluster may or may not have dissolved by this time. We consider two types of gas expulsion modes, i.e. "fast" ($\tau_g$) and "moderate" (5$\tau_g$). $\tau_g$ is the gas depletion time in Eq.\ref{eq:mdecay}. "yes" or "no" indicate whether spatial separations or relative velocities between embedded clusters are accounted for in the simulations. Both "NGC1893-fast-vd-d0.5" and "NGC1893-fast-vd" incorporate spatial separations between embedded clusters, but the separation of the former is half of the latter.
}
\label{tab1}
\begin{tabular}{ccccc}
\hline
Cases	&	Gas expulsion	&	Spatial separation	&	Relative velocity	&	Terminated time (Myr)	\\
NGC1893-fast	&	fast	&	no	&	no	&	100	\\
NGC1893-fast-vd	&	fast	&	yes	&	yes	&	100	\\
NGC1893-fast-vd-v0	&	fast	&	yes	&	no	&	100	\\
NGC1893-fast-vd-d0.5	&	fast	&	yes,  * 0.5	&	yes	&	100	\\
NGC1893-moderate-vd	&	moderate	&	yes	&	yes	&	100	\\
NGC6334-fast	&	fast	&	no	&	no	&	614.25	\\
NGC6334-fast-vd	&	fast	&	yes	&	yes	&	463.75	\\
NGC6334-fast-vd-v0	&	fast	&	yes	&	no	&	100	\\
NGC6334-moderate-vd	&	moderate	&	yes	&	yes	&	259.25	\\
Carina-fast	&	fast	&	no	&	no	&	556.5	\\
Carina-fast-vd	&	fast	&	yes	&	yes	&	324.5	\\
\hline
\label{case}
\end{tabular}
\end{table*}

\subsubsection{N-body simulation}\label{recipe}
The parameters for the simulation are summarized in previous works, i.e.
\citet{Kroupa2001-321,
Baumgardt2007-380,
Banerjee2012-746,
Banerjee2013-764,
Banerjee2014-787,
Banerjee2015-447,
Oh2015-805,
Oh2016-590,
Banerjee2017-597,
Brinkmann2017-600,
Oh2018-481,
Wang2019-484,
Pavlik2019-626,
Dinnbier2022-660,Zhou2024rm,Zhou2024sub}.
Previous simulations have demonstrated the effectiveness and appropriateness of the parameter settings (see below). The impact of varying parameter settings on simulation outcomes, as well as the exploration of multi-dimensional parameter space, are also discussed in the referenced literature. In this study, we primarily apply the established simulation framework to interpret observational data.
For a detailed description and discussion of the initial conditions and parameters of the simulation can refer to \citet{Zhou2024sub}. Here, we provide only a brief overview.

The initial density profile of all clusters is the Plummer profile \citep{Aarseth1974-37, HeggieHut2003, Kroupa2008-760}. The half-mass radius $r_{h}$ of the cluster is given by the $r_{\rm h}-M_{\rm ecl}$ relation \citep{Marks2012-543}. All clusters are fully mass segregated ($S$=1), no fractalization, and in the state of virial equilibrium ($Q$=0.5). $S$ and $Q$ are the degree of mass segregation and the virial ratio of the cluster, respectively. More details can be found in \citet{Kupper2011-417} and the user manual for the \texttt{McLuster} code. Discussions regarding these parameters can be found in \citet{Banerjee2013-764,Zhou2024rm}.
The initial mass functions (IMFs) of the clusters are chosen to be canonical \citep{Kroupa2001-322} with the most massive star following the $m_{\rm max}-M_{\rm ecl}$ relation  of \citet{Weidner2013-434}. We assume the clusters to be at solar metallicity, i.e. $Z=0.02$ \citep{von2016ApJ...816...13V}. 
The clusters traverse circular orbits within the Galaxy, positioned at a Galactocentric distance of 8.5 kpc, moving at a speed of 220 km s$^{-1}$.
The initial binary setup follows the method described in \citet{Wang2019-484}.
All stars are initially in binaries, i.e. $f_{\rm b}$=1, where $f_{\rm b}$ is the primordial binary fraction.

Modeling gas removal from embedded clusters poses significant challenges due to the complexities of radiation hydrodynamical processes, which involve intricate and uncertain physical mechanisms. To simplify the approach, we simulate the primary dynamical effects of gas expulsion by applying a diluting, spherically symmetric external gravitational potential to a model cluster, following the methods of \citet{Kroupa2001-321} and \citet{Banerjee2013-764}. This analytical method receives partial validation from \citet{Geyer2001-323}, who conducted comparison simulations using smoothed particle hydrodynamics to represent the gas. Additionally, hydrodynamics+N-body simulations by \citet{Farias2024-527} affirm that the exponential decay model described in Eq.\ref{eq:mdecay} effectively captures the gas removal process driven by radiation and wind feedback. 
Specifically, we adopt the potential of a spherically symmetric, time-dependent mass distribution
\begin{eqnarray}
M_g(t)=& M_g(0) & t \leq \tau_d,\nonumber\\
M_g(t)=& M_g(0)\exp{\left(-\frac{(t-\tau_d)}{\tau_g}\right)} & t > \tau_d.
\label{eq:mdecay}
\end{eqnarray}
Here, $M_g(t)$ is the total mass in gas which is spatially distributed with the Plummer density distribution \citep{Kroupa2008-760} and starts depleting with timescale $\tau_g$ after a delay of $\tau_d$. The Plummer radius of the gas distribution is kept time-invariant \citep{Kroupa2001-321,Zhou2024sub}.
We use an average velocity of the outflowing gas of $v_g\approx10$ km s$^{-1}$, which is the typical sound-speed in an HII region. This gives
$\tau_g=r_h(0)/v_g$,
where $r_h(0)$ is the initial half-mass radius of the cluster. As for the delay-time, we take the representative value of $\tau_d\approx0.6$ Myr
\citep{Kroupa2001-321}, this being about the life-time of the ultra-compact HII region. 
In this work, we assume a star formation efficiency SFE $\approx$ 0.33 as a benchmark \citep{Lada2003-41,Megeath2016-151,Zhou2024PASP-1,Zhou2024PASP-2}, i.e. $M_{g}(0)$ = 2$M_{\rm ecl}(0)$.

The \texttt{McLuster} program \citep{Kupper2011-417} was used to set the initial configurations. 
The dynamical evolution of the model clusters was computed using the state-of-the-art \texttt{PeTar} code \citep{Wang2020-497}. 
\texttt{PeTar} employs well-tested analytical single and binary stellar evolution recipes (SSE/BSE)
\citep{Hurley2000-315,Hurley2002-329,Giacobbo2018-474,Tanikawa2020-495,Banerjee2020-639}.

\subsubsection{Initial conditions}

\begin{figure*}
\centering
\includegraphics[width=1\textwidth]{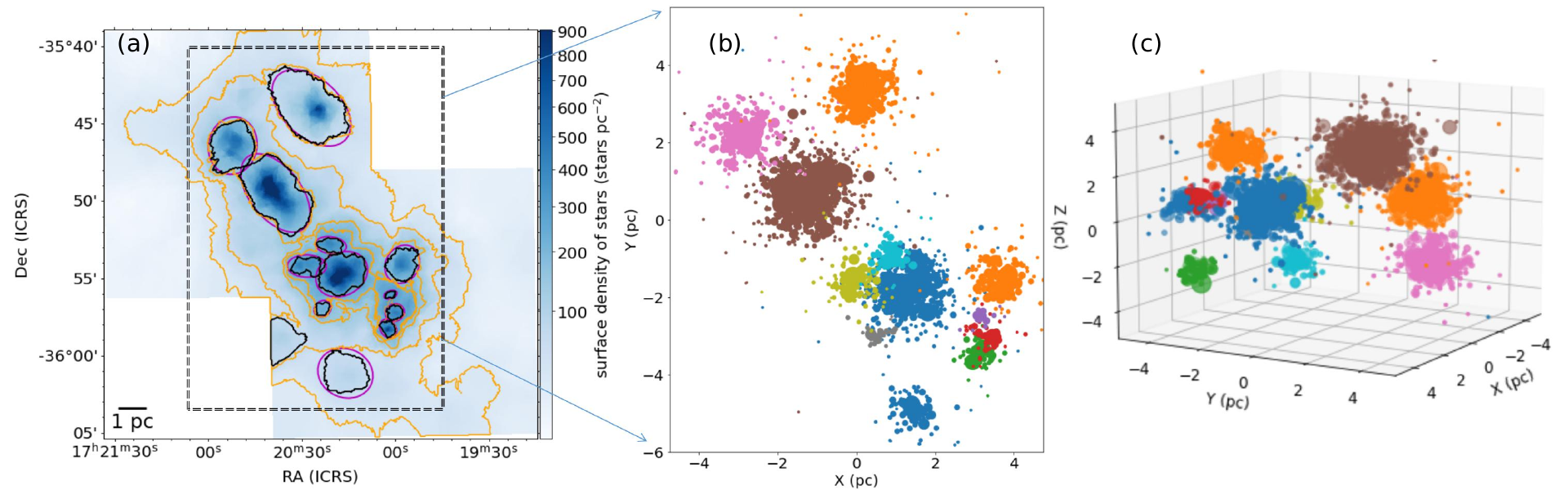}
\caption{The initial conditions of the simulations drawn from the observations of NGC 6334.
(a) Background is the surface density of stars. Black contours show the masks of embedded clusters identified by the dendrogram algorithm in \citet{Zhou2024-688}. Magenta ellipses are the approximate ellipses that encompass the embedded clusters;
(b) The collection of all simulated embedded clusters which are used to create an initial configuration of the embedded cluster complex similar to the observation; (c) Considering spatial separations between embedded clusters based on panel (b). In panels (b) and (c), the size of the dot scales with the stellar mass using an arbitrary scaling.}
\label{example}
\end{figure*}

\begin{figure*}
\centering
\includegraphics[width=1\textwidth]{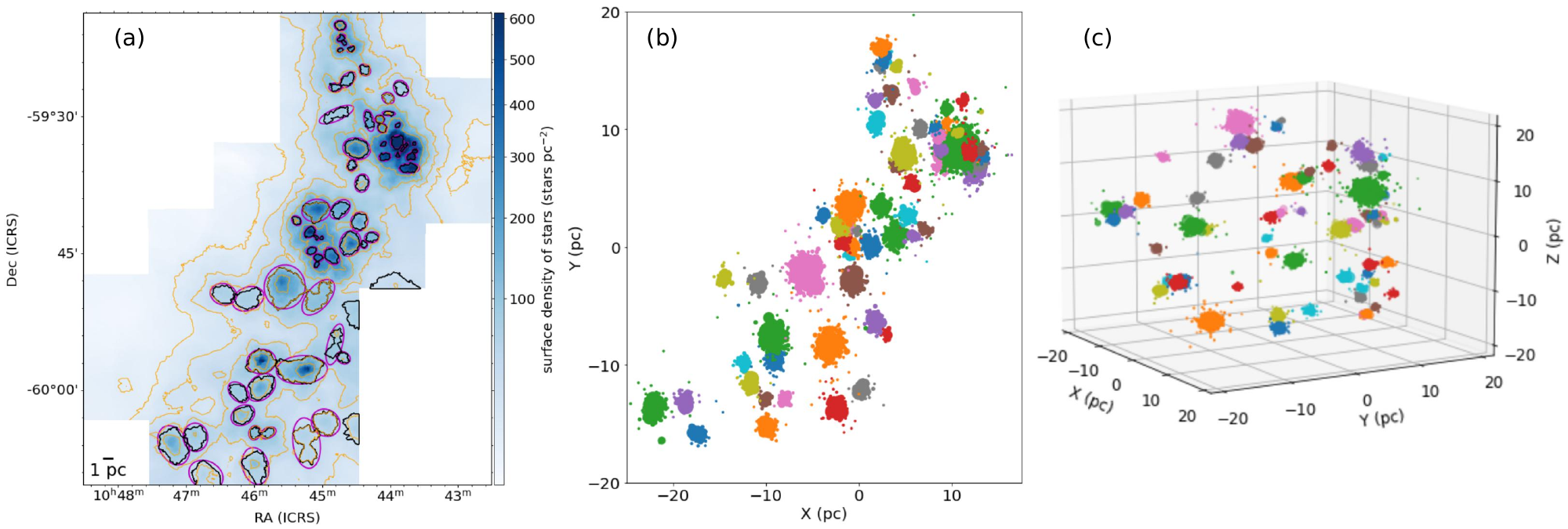}
\caption{Same as Fig.~\ref{example}, but for Carina.}
\label{example1}
\end{figure*}

\begin{figure*}
\centering
\includegraphics[width=1\textwidth]{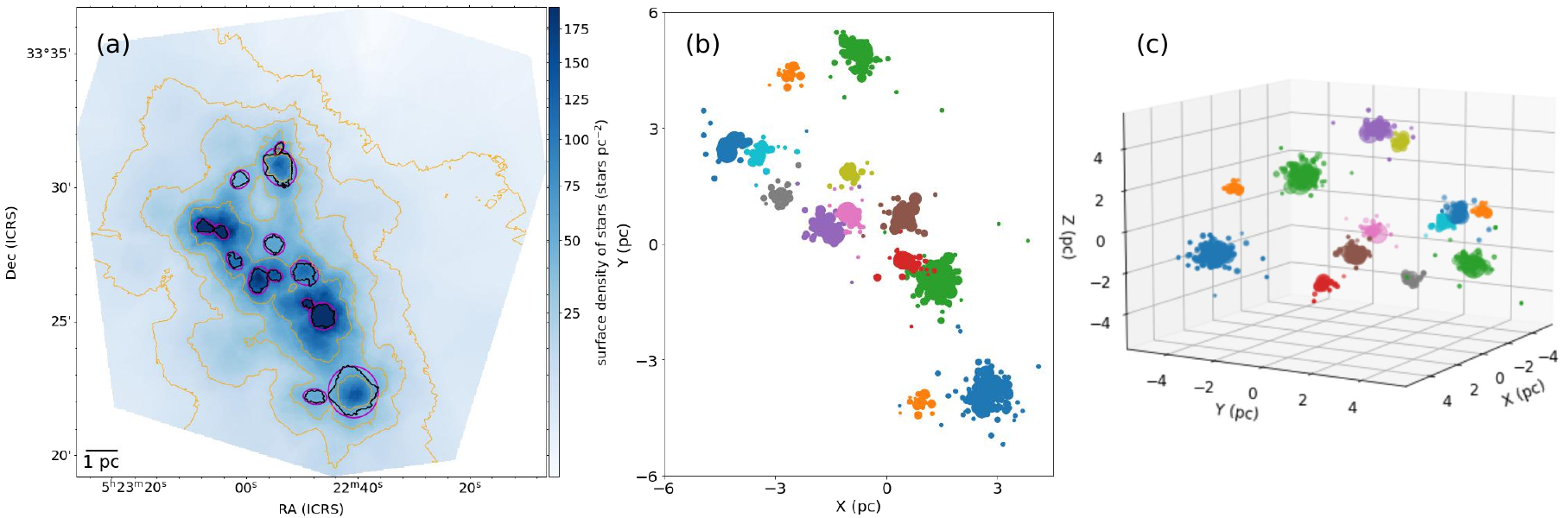}
\caption{Same as Fig.~\ref{example}, but for NGC 1893.}
\label{example2}
\end{figure*}

For the 17 MSFRs investigated in the MYStIX project, nine of them are subcluster complexes. As shown in \citet{Kuhn2014-787}, some of these nine MSFRs exhibit similar morphologies. Finally, we selected NGC 1893, NGC 6334 and the Carina nebula as representatives. In these three MSFRs, the spatial distribution of their subclusters varies, particularly with significant differences in the total mass of the subclusters, which helps to fit the observed physical parameters of open clusters.
We have identified the subclusters in these three MSFRs using the dendrogram algorithm according to the surface density distributions of stars in \citet{Zhou2024-688}, and calculated the physical parameters of the subclusters. Considering that
the mass-radius relation of embedded clusters can be well fitted by the $\approx$1 Myr expanding line in \citet{Zhou2024rm}, 
we simulated the evolution of the subclusters in each MSFR for the first 1 Myr using the recipe described above. The initial masses of the subclusters strictly follow the observed values, and each subcluster is simulated individually.
Then, all subclusters were collected  together to create an initial configuration similar to the observation, as shown in Fig.~\ref{example}. The separations between subclusters in Fig.~\ref{example}(b) strictly adhere to the observed values. 

\begin{figure}
\centering
\includegraphics[width=0.45\textwidth]{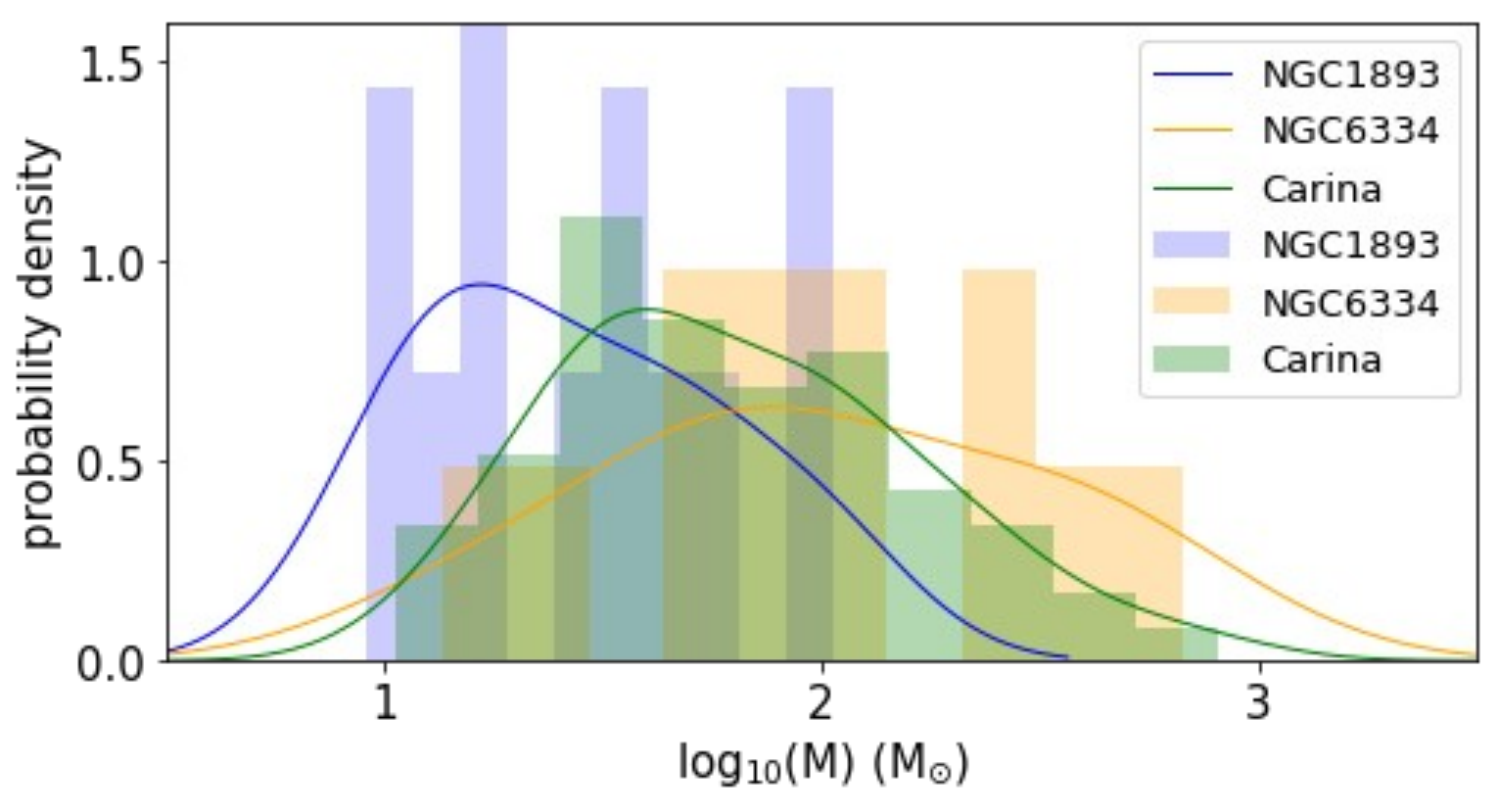}
\caption{Mass distribution of embedded clusters in three MSFRs.}
\label{mass}
\end{figure}

\begin{figure*}
\centering
\includegraphics[width=0.95\textwidth]{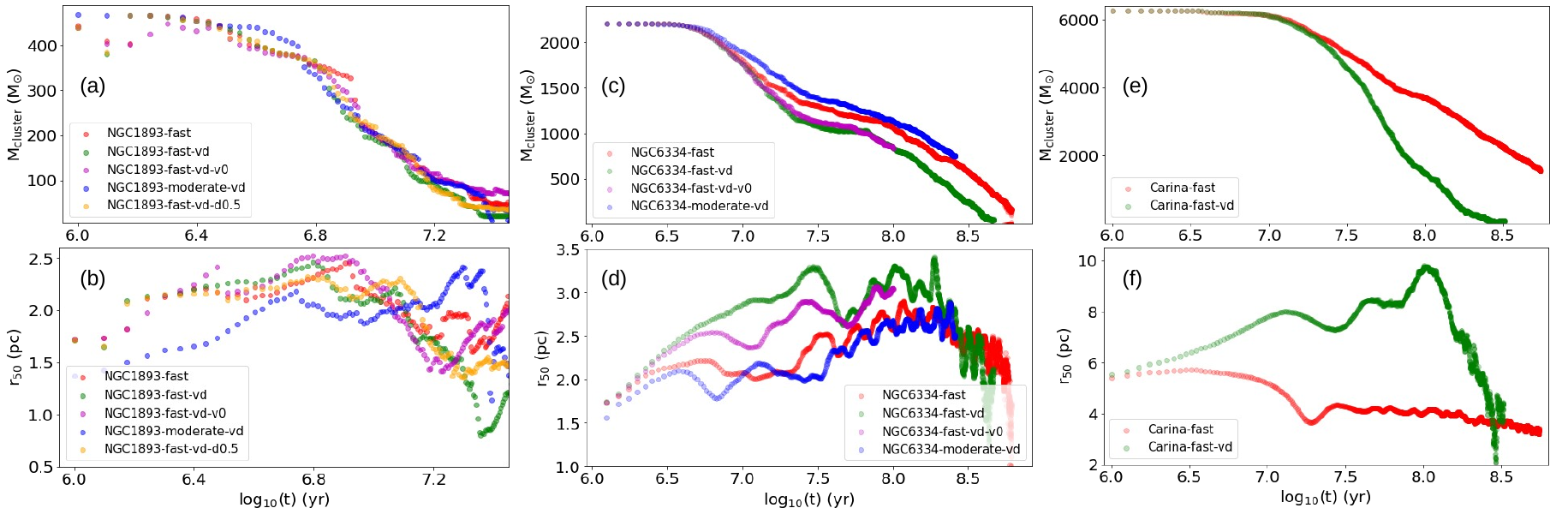}
\caption{Mass and radius evolution of the embedded cluster complexes over time in different cases listed in Table.\ref{case}. r$_{50}$ is the radius containing 50\% of the members within the tidal radius of the cluster.}
\label{Mrt}
\end{figure*}

\subsubsection{Gas expulsion}

The physical processes involved in stage1, stage2 and part of stage3 lead to the current spatial and mass distributions of clumps and star clusters in MSFRs that we observe today. In this work, our simulations start directly from the current state of MSFRs. Star clusters have separated from the gas (see Appendix.A of \citet{Zhou2024sub}) and their subsequent evolution is primarily driven by the dynamical interactions within and between star clusters, such that we can focus solely on the N-body dynamics. However, the evolution of MSFRs is a continuous process, and events that occurred in the past still have an impact on the present. 
The main physical process related to the gas in 
stage3 is the initial gas expulsion due to feedback, which also provides the initial conditions for the subsequent evolution of the cluster in N-body simulations. The initially embedded clusters are gravitationally bound by the surrounding gas. After the gas expulsion, the clusters will expand and subsequently interact and coalescence. 
Modeling the gas expulsion process is crucial for understanding the current physical state of the star clusters.

As discussed in \citet{Zhou2024PASP-2} and \citet{Zhou2024sub}, more massive clumps are capable of forming more massive clusters, which results in stronger feedback and higher gas expulsion velocity. Consequently, we expect correlations between the strength of feedback, the mass of the clumps (or clusters), and the gas expulsion velocity. Low-mass clusters should experience slower gas expulsion compared to high-mass clusters. The star formation efficiency (SFE) determines the total amount of the remaining gas, which also influences the timescale of gas expulsion. Additionally, the complex, non-spherical spatial distribution of gas can further alter this timescale. 
In summary, the uncertainties in these parameters (gas expulsion velocity, SFE and spatial distribution of gas) can ultimately be factored into the gas expulsion timescale. Moreover, the amount of residual gas not only influences the expulsion process but also determines the gravitational potential of the gas. As discussed in \citet{Zhou2024sub}, a lower SFE leads to more residual gas, which drives more intense expansion and greater mass loss in the cluster after gas expulsion. Essentially, a lower SFE is equivalent to a shorter gas expulsion timescale.
Thus, apart from the fast gas expulsion with the gas depletion time of $\tau_{\rm g}$, we also simulated a moderate gas expulsion with the gas depletion time of 5$\tau_{\rm g}$ for each embedded cluster.

\subsubsection{Spatial distribution and velocity difference}

Currently, we assume all embedded clusters are initially at rest and located on the same plane. A more realistic scenario is that clusters have relative velocities and line-of-sight spatial separations. To take these two factors into account, we approximate the molecular cloud containing the clusters as an ellipsoid. To derive the lengths of the three axes of the ellipsoid (semi-major axis, $a$, intermediate axis, $b$, semi-minor axis, $c$), we fit the coordinates of the clusters located at the periphery to determine $a$ and $b$. Given that the real cloud shape should be more sheet-like instead of spherical \citep{Shetty2006-647,Inutsuka2015-580,Arzoumanian2018-70,Kohno2021-73,Arzoumanian2022-660,Rezaei2022-930,Zhou2023-519,Zhou2023-676,Clarke2023-519,Ganguly2023-525}, we assume $c=b$.
Then we randomly distribute clusters along the Z-axis (line-of-sight). As shown in Fig.~\ref{example}, Fig.~\ref{example1}
and Fig.~\ref{example2}, the spatial distribution of clusters is nearly spherical. Therefore, we have overestimated the spacing between clusters.

For molecular clouds in the catalog of \citet{Miville2017-834} across the Galactic plane, the most probable cloud size is $\approx$30 pc, and the most probable value of the velocity dispersion is $\approx$1.95 km s$^{-1}$ on the scale of clouds. 
In our case, the sizes of the MSFRs are $\approx$5 pc for NGC 6334 and NGC 1893, and $\approx$20 pc for the Carina MSFR.
As a conservative estimate, we take the velocity dispersion of the original molecular clouds in the three MSFRs as 2 km s$^{-1}$ and assume that the clusters inherit the velocity dispersion of the clouds. The system's center has a velocity of 0 km s$^{-1}$, and the velocity of the outermost cluster is 2 km s$^{-1}$. The velocities of other clusters are distributed according to the Larson relation, i.e. $v \propto L^{0.5}$, where $L$ is the distance to the system's center. 

\subsubsection{Simulation results analysis}\label{analysis}

The main objective of this study is to assess whether the post-gas expulsion coalescence of embedded clusters can explain the observed physical parameters of open clusters.
Figure~\ref{ngc1893-fast}, Fig.~\ref{ngc6334-fast}, Fig.~\ref{ngc6334-fast-vd}, Fig.~\ref{ngc6334-fast-vd-3D}, Fig.~\ref{Carina-fast} and Fig.~\ref{Carina-fast-vd} present the evolution of the embedded cluster complexes over time. 
In the simulations, before 1 Myr, the evolution of each cluster is approximately independent. After further expansion, they start to interact upon contact, leading to subsequent mutual influence in their evolution.
The spatial distribution, relative velocities, mass distribution, and gas expulsion modes of embedded clusters will affect the dynamics of their coalescence as well as the stability of the coalesced products.
We simulated all possible cases listed in Table.~\ref{case} for NGC 1893.
The embedded clusters in NGC 1893 have low mass (see Fig.\ref{mass}) and are few in number (13). They will quickly disperse after expansion regardless of the initial conditions, as shown in Fig.~\ref{Mrt}(a) and Fig.~\ref{ngc1893-fast}. There may not be a real coalescence between them, but rather the formation of a loose association. 

For NGC 6334, we simulated four cases, i.e. "fast", "fast-vd", "fast-vd-v0" and "moderate-vd". Overall, they produce similar results, but there are also differences. 
The line-of-sight spatial separations and 
relative velocities between embedded clusters ("fast-vd") lead to a larger radius and faster mass loss of the coalesced cluster, but this effect can be counteracted by a slower gas expulsion of the embedded clusters ("moderate-vd"). Therefore, there is a degeneracy between the line-of-sight spatial separations and the relative velocities between embedded clusters and the gas expulsion mode of embedded clusters. Thus, "moderate-vd" and "fast" give comparable simulation results, as shown in Fig.~\ref{Mrt}(c) and (d). Comparing "fast-vd" and "fast-vd-v0", we can see that the line-of-sight spatial separations are more important than the relative velocities in determining the mass loss and radius of the coalesced cluster.

NGC 6334 and NGC 1893 both have a small number of embedded clusters, but NGC 6334 contains more massive embedded clusters (see Fig.~\ref{mass}). 
By comparing Fig.~\ref{ngc1893-fast} and Fig.~\ref{ngc6334-fast}, it is evident that the mass of embedded clusters plays a critical role in the coalescence process. Massive embedded clusters provide a more stable gravitational core, significantly aiding the coalescence and enabling the resulting cluster to remain stable within the Galactic tidal field. Strong gravitational forces between the embedded cluster complexes are essential to counteract the system's expansion and the Galactic tidal influence, ensuring the coalescence process continues. This strong local gravitational field is predominantly generated by the massive embedded clusters. In Fig.~\ref{ngc6334-fast}, we can see the coalescence between embedded clusters in NGC 6334.

Most of the embedded clusters in the Carina complex are low-mass ($<$100 M$_{\odot}$, see Fig.~\ref{mass}), but they are numerous and densely distributed. Additionally, the Carina also contains a few massive embedded clusters ($>$300 M$_{\odot}$). Therefore, even though the embedded clusters in Carina were initially given large line-of-sight spatial separations (the maximum $\sim$40 pc), they are still able to coalesce after expansion, which is significantly different from the case of NGC 1893. However, this coalescence complex is not stable enough and gradually dissipates in the Galactic tidal field, as shown in Fig.~\ref{Carina-fast-vd}. The large line-of-sight spatial separations lead to insufficient coalescence between embedded clusters, preventing the formation of a stable core. In contrast, since there is no line-of-sight spatial separation in the case of "Carina-fast", the embedded clusters are able to sufficiently coalesce, forming a stable cluster core that can persist for over 300 Myr, as shown in Fig.~\ref{Carina-fast}.


\subsection{Comparison with observations}\label{compare}

\begin{figure}
\centering
\includegraphics[width=0.45\textwidth]{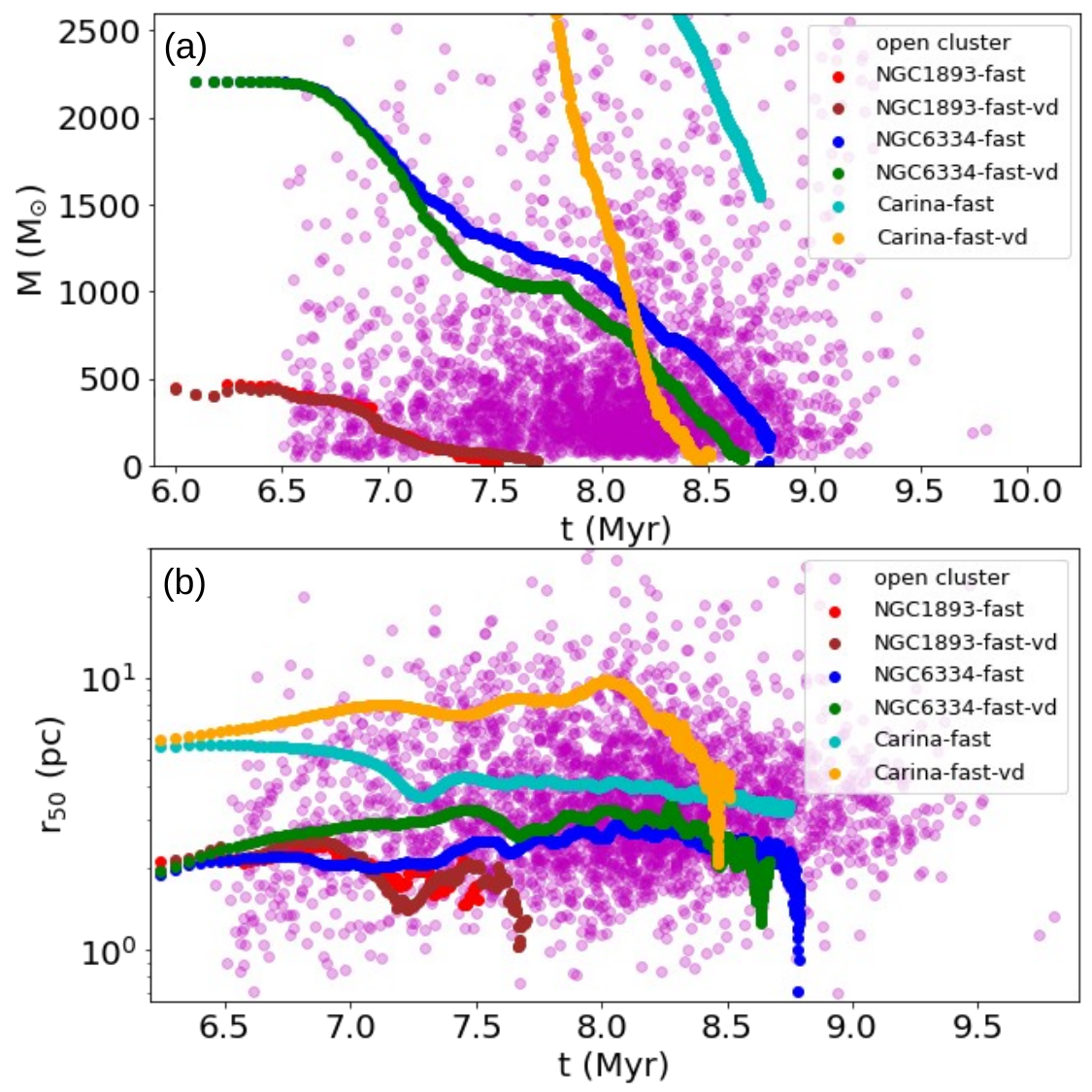}
\caption{Fitting the physical parameters of open clusters in \citet{Hunt2024-686} using the coalescence simulations. r$_{50}$ is the radius containing 50\% of the members within the tidal radius of the open cluster. $M$ is the mass of the open cluster.}
\label{fit}
\end{figure}

The primary goal of this work is to evaluate whether the post-gas expulsion coalescence of embedded clusters can account for the observed physical parameters of open clusters.
Therefore, we only consider two extreme cases, i.e. "fast" and "fast-vd". The simulation results from other parameter choices should lie in between these two extremes, as shown in Fig.~\ref{Mrt}.

As presented in \citet{Zhou2024sub},
to explain the mass and radius distributions of the observed open clusters, initial embedded clusters with masses larger than 3000 M$_{\odot}$ are necessary. 
However, the upper limit of the embedded cluster sample collected in \citet{Zhou2024sub} is less than 1000 M$_{\odot}$, and only few ATLASGAL clumps have a mass larger than 3000 M$_{\odot}$.
The current mass distribution of clumps in the Milky Way suggests that the evolutionary process, where a single clump develops into an embedded cluster and eventually into an open cluster, cannot account for the observed open clusters with relatively old ages and large masses.
However, each molecular cloud typically contains multiple clumps, which are the local sites of star formation within the molecular cloud and the precursors of embedded clusters. As presented above, the formed embedded clusters can coalesce after expanding. 
If we require that open clusters evolve directly from single embedded clusters, then the embedded cluster that serve as the precursor must have a sufficiently high mass to account for the observed physical parameters of open clusters. However, the currently observed clump masses make it difficult to directly form massive embedded clusters. Nevertheless, a composite of multiple low-mass embedded clusters could, through coalescence, cover the parameter space observed in open clusters, as shown in Fig.~\ref{fit}, where the mass and radius of the clusters are displayed as a function of their age. In the post-gas expulsion coalescence scenario, massive embedded clusters are not necessary. This is consistent with the current mass distribution of clumps in the Milky Way.
We note that, as presented in \citet{Zhou2024sub}, the evolution of individual embedded clusters can also account for part of the observed parameter space of open clusters. 

\section{Discussion}\label{discuss}

The structures observed in the surface density maps of stars from the MYStIX project represent the projected distributions of stars within star-forming regions, leading to possible overlaps of physically separate star groups on the map. Additionally, some of the small dense embedded clusters might
not be visible in the surface density maps or not be identified by the dendrogram algorithm.
Moreover, we may underestimate the mass of the embedded clusters, due to assuming an average mass of 0.5 M$_{\odot}$ for all stars in each embedded cluster. However, in this work, we simply aim to obtain a reasonable spatial distribution of embedded clusters in MSFRs based on observations. This work does not seek to accurately represent the evolution of embedded clusters within NGC 6334, NGC 1893 and the Carina. 

As described in Sec.~\ref{compare}, based on the current mass distribution of clumps and embedded clusters in the Milky Way, combining the evolution of individual embedded clusters and the coalescence of multiple embedded clusters can encompass the parameter space of the observed open clusters. This work focuses on the coalescence of the formed embedded clusters after expansion. Actually, there is now extensive literature arguing that star clusters form through the mergers of subclusters, both from simulations and observations.
The N-body simulations of \citet{Fujii2012-753} revealed that clusters formed through the hierarchical merging of multiple subclusters can account for the characteristics (such as mass distribution, mass segregation, runaway stars, massive stars, and massive binaries) observed in young dense clusters like NGC 3603 and Westerlund 1 and 2 in the Milky Way, as well as R136 in the Large Magellanic Cloud (LMC). 
The idea that runaway stars originate from subcluster mergers is also explored in \citet{Polak2024arXiv240512286P}.
\citet{Sills2018-477} conducted simulations of young, embedded star clusters, modeling the dynamical evolution of stars and gas simultaneously, with initial conditions derived directly from observations in the MYStIX project.
They found that initially highly sub-structured systems rapidly evolved into spherical, monolithic, and smooth distributions in both spatial and velocity distributions, similar to those shown in Fig.~\ref{ngc6334-fast}, Fig.~\ref{ngc6334-fast-vd}, Fig.~\ref{ngc6334-fast-vd-3D}, Fig.~\ref{Carina-fast} and Fig.~\ref{Carina-fast-vd}. Although the simulations of \citet{Sills2018-477} and those presented in this work differ in some aspects, the results are similar, indicating that the coalescence of subclusters is not sensitive to the initial conditions.

On the observational side, 
\citet{Gennaro2011-412} proposed that the elongation and mass segregation observed in Westerlund 1 could be the result of a merger of multiple subclusters that formed nearly coevally within the parent giant molecular cloud (GMC).
\citet{Sabbi2012-754} reported
the identification of two distinct stellar populations in the core of the giant H II region 30 Doradus in the LMC. The differences in morphology and age between the two subclusters suggest that an ongoing merger may be taking place within the core of 30 Doradus.
\citet{Della2023-674} identified a massive, substructured stellar system (called "LISCA II") in the Galactic Perseus complex that is likely undergoing hierarchical cluster assembly. This system consists of nine star clusters. They observed a dominant contraction pattern toward the center of the system, primarily driven by the external regions of the clusters' complex, along with a milder central expansion. These observations align with expectations for the early evolutionary phases of stellar systems forming into a coherent massive structure, as predicted by their N-body models. In Fig.~\ref{Carina-fast}, we can see some similar physical processes described above. 
Another sub-structured stellar system (called "LISCA I") was identified in \citet{Dalessandro2021-909}, which is populated by seven co-moving clusters with signs of potential interactions. After comparing with direct N-body simulation, the authors suggest that LISCA I could be in an intermediate stage of cluster assembly, which may eventually evolve into a more massive stellar system.
Although the subcluster merger simulations in \citet{Dalessandro2021-909} used very different initial conditions, the hierarchical cluster formation path presented in their work (Figure 7 of \citet{Dalessandro2021-909}) is comparable with Fig.~\ref{Carina-fast} and Fig.~\ref{Carina-fast-vd} in this work. Thus, within a reasonable range of parameters, the coalescence of subclusters is a robust process, not sensitive to the initial conditions.

\section{Conclusion}\label{conc}

Based on the spatial and mass distributions of embedded clusters in three massive star-forming regions (MSFRs) from the observations (NGC 1893, NGC 6334, and the Carina Nebula), 
we simulated the evolution of the embedded clusters in each MSFR to $\approx$1 Myr, and then collected them together to create an initial configuration of the embedded cluster complex similar to that in the observation of each MSFR. 
In the simulations, before $\approx$1 Myr, the evolution of each embedded cluster is approximately independent. After further expansion, they start to interact upon contact, leading to subsequent mutual influence in their evolution.
The spatial distribution, relative velocities, mass distribution and gas expulsion modes of embedded clusters will affect the dynamics of their coalescence process as well as the stability of the coalesced products. The large line-of-sight spatial separations lead to insufficient coalescence between embedded clusters, preventing the formation of a stable core. 
Greater spatial separations and higher relative velocities between embedded clusters lead to larger radius and faster mass loss of the coalesced star cluster, but this effect can be counteracted by a slower gas expulsion of the embedded clusters. 
By comparing with other independent works, we conclude that, within a reasonable range of parameters, the coalescence of embedded clusters is a robust process, not sensitive to the initial conditions.

The embedded clusters in NGC 1893 have low masses and are few in numbers. They will quickly disperse after expansion regardless of the initial conditions of the simulations. There may not be a real coalescence between them, but rather the formation of a loose association. 
NGC 6334 and NGC 1893 both have a small number of embedded clusters, but NGC 6334 contains more massive embedded clusters, leading to the coalescence of embedded clusters in NGC 6334. Thus, the embedded cluster mass plays a critical role in the coalescence process. Massive embedded clusters provide a more stable gravitational core, significantly aiding the coalescence and enabling the resulting cluster to counteract the system's expansion and the Galactic tidal influence. By comparing the Carina and NGC 1893 MSFRs, it seems that a larger number of embedded clusters can also facilitate their coalescence.

The currently observed clumps in the Milky Way are not massive enough to directly form massive embedded clusters.
However, a composite of multiple low-mass embedded clusters could, through coalescence, cover the parameter space of the observed open clusters in the Milky Way. In the post-gas expulsion coalescence scenario, open clusters do not necessarily require massive embedded clusters as precursors.

\section*{Acknowledgements}
We would like to thank the referee for the detailed comments and suggestions, which have helped to improve and clarify this work.

\section{Data availability}
All the data used in this work are available from the first author.

\bibliography{ref}
\bibliographystyle{aasjournal}


\begin{appendix}
\twocolumn

\section{Snapshots of N-body simulations}

\begin{figure}
\centering
\includegraphics[width=0.45\textwidth]{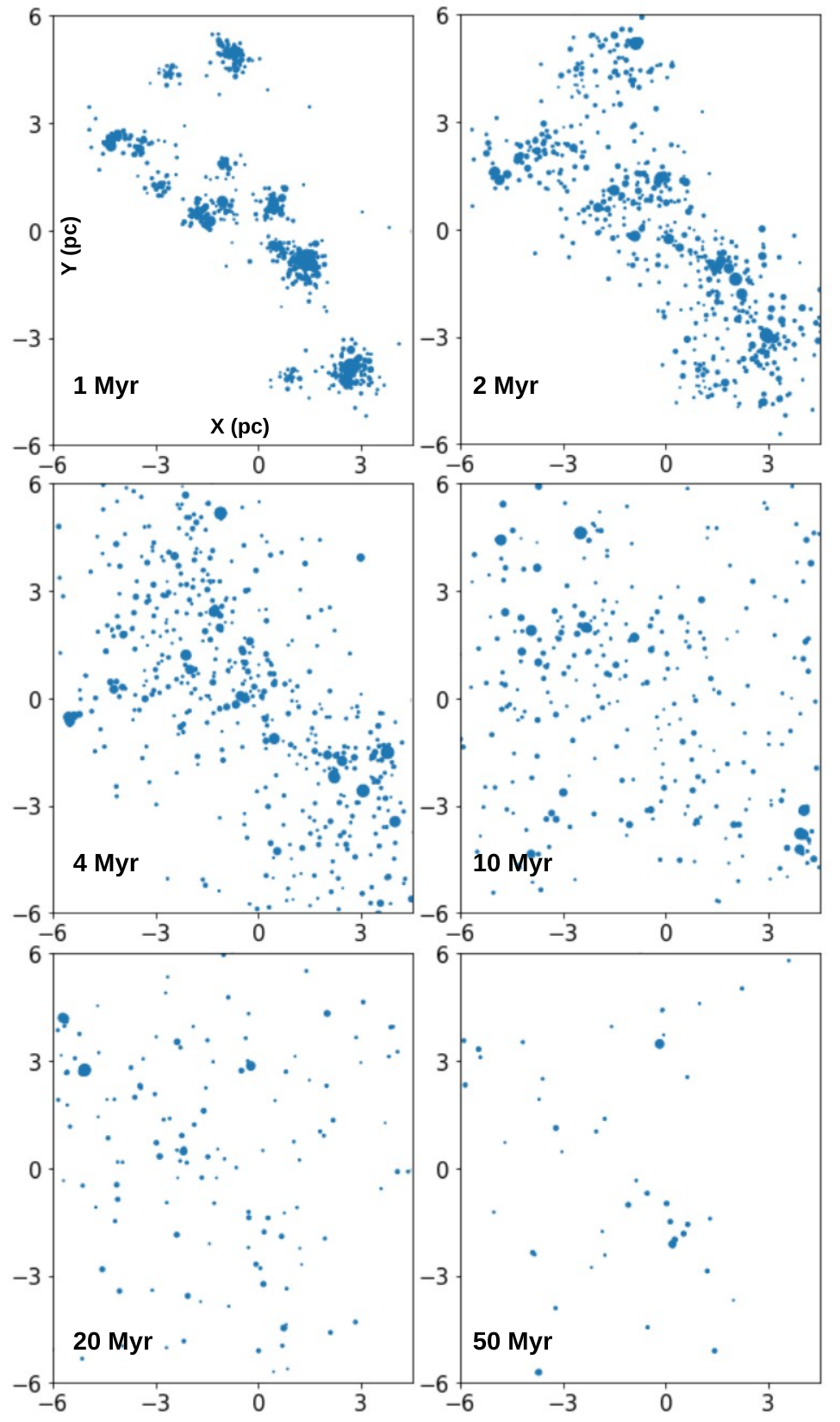}
\caption{The snapshots on the XY plane in the case of "NGC1893-fast" starting from Fig.~\ref{example2}(b). The size of the points in the map represents the stellar mass.}
\label{ngc1893-fast}
\end{figure}

\begin{figure}
\centering
\includegraphics[width=0.45\textwidth]{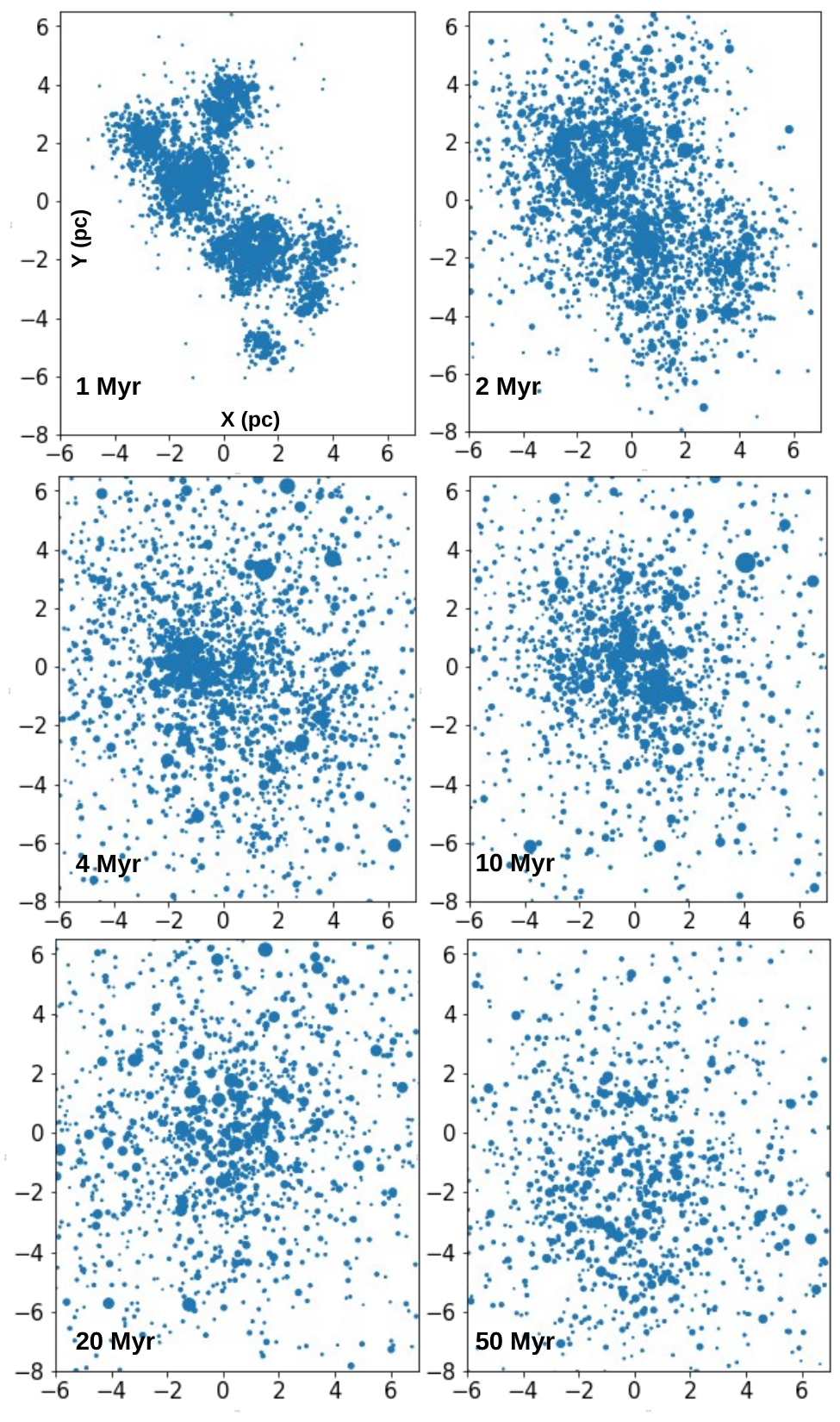}
\caption{The snapshots on the XY plane in the case of "NGC6334-fast" starting from Fig.~\ref{example}(b). The size of the points in the map represents the stellar mass.}
\label{ngc6334-fast}
\end{figure}

\begin{figure}
\centering
\includegraphics[width=0.45\textwidth]{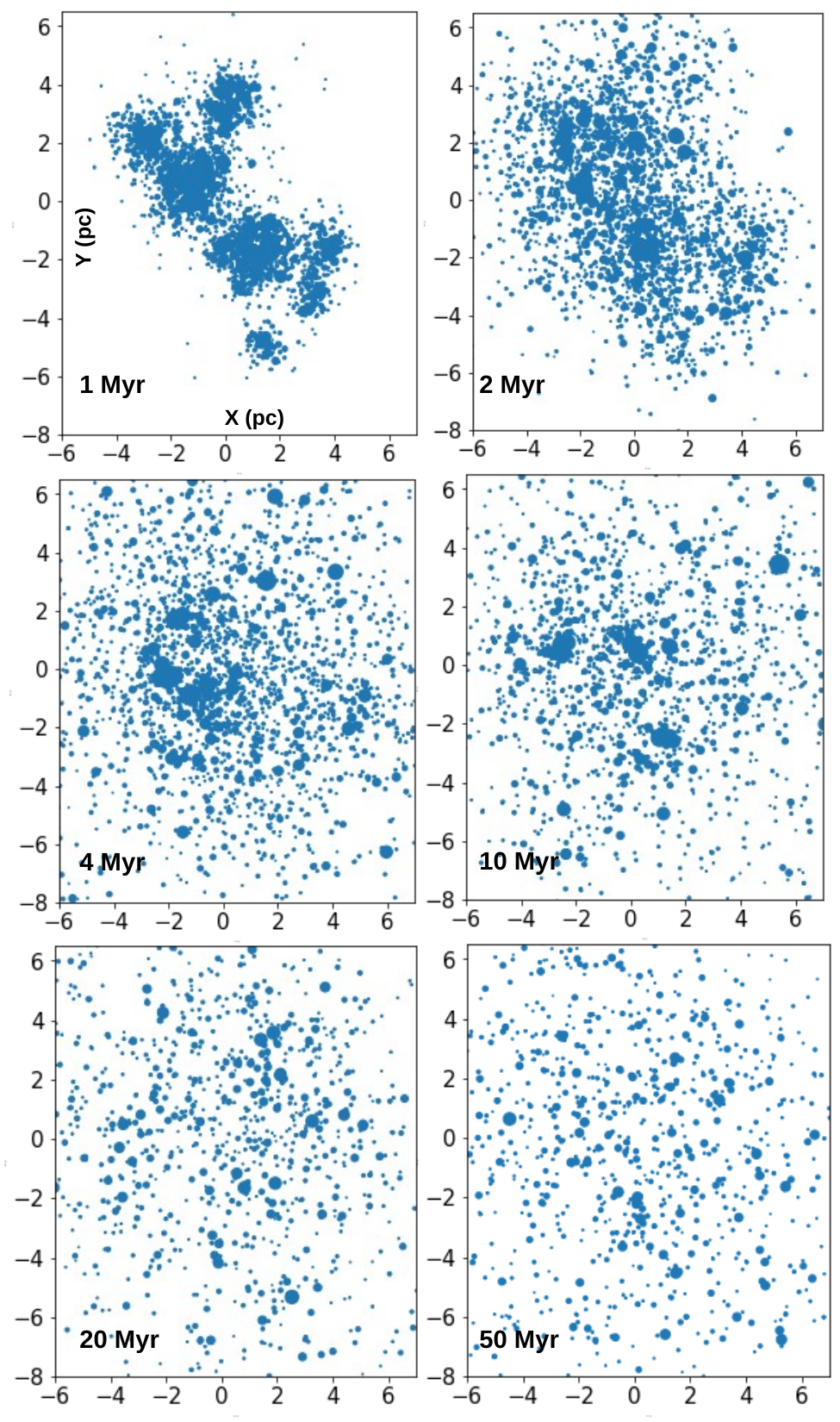}
\caption{The snapshots on the XY plane in the case of "NGC6334-fast-vd" starting from Fig.~\ref{example}(c). The size of the points in the map represents the stellar mass.}
\label{ngc6334-fast-vd}
\end{figure}

\begin{figure}
\centering
\includegraphics[width=0.48\textwidth]{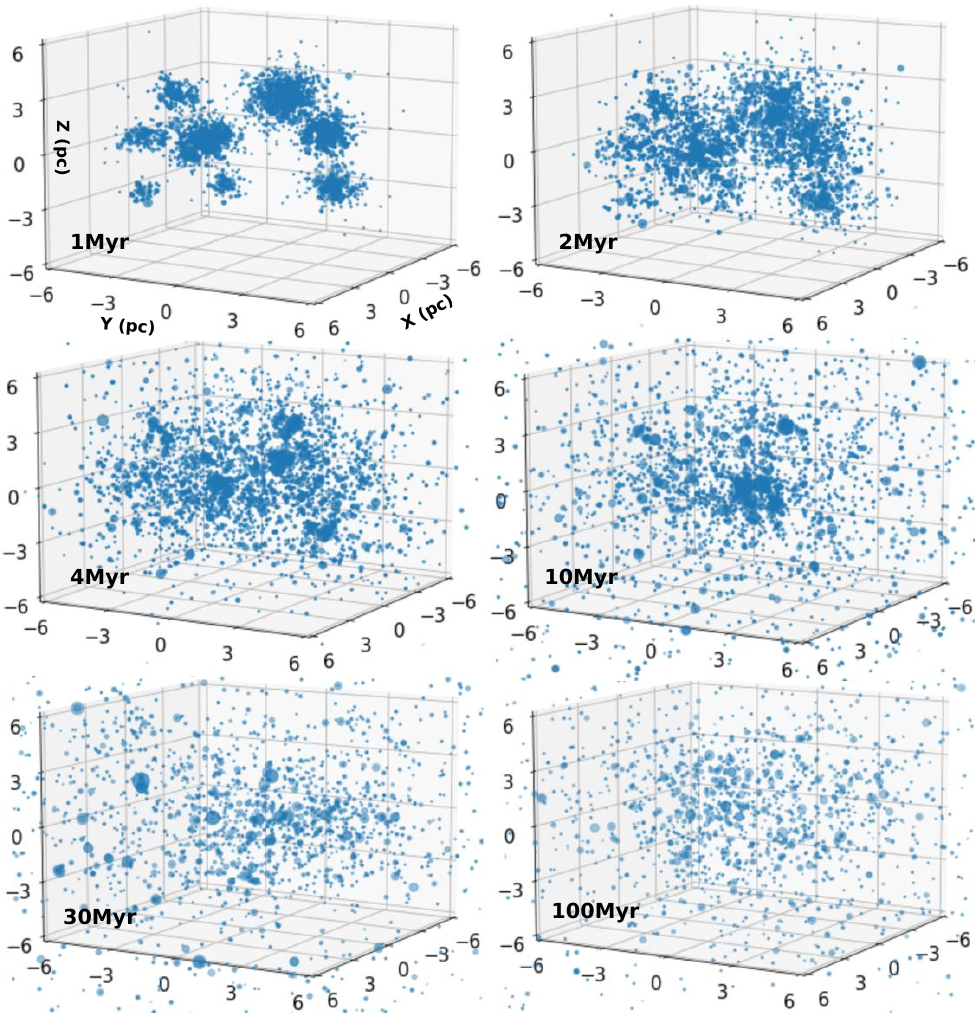}
\caption{The snapshots in the 3D space (XYZ) in the case of "NGC6334-fast-vd" starting from Fig.\ref{example}(c). The size of the points represents the stellar mass.}
\label{ngc6334-fast-vd-3D}
\end{figure}

\begin{figure}
\centering
\includegraphics[width=0.45\textwidth]{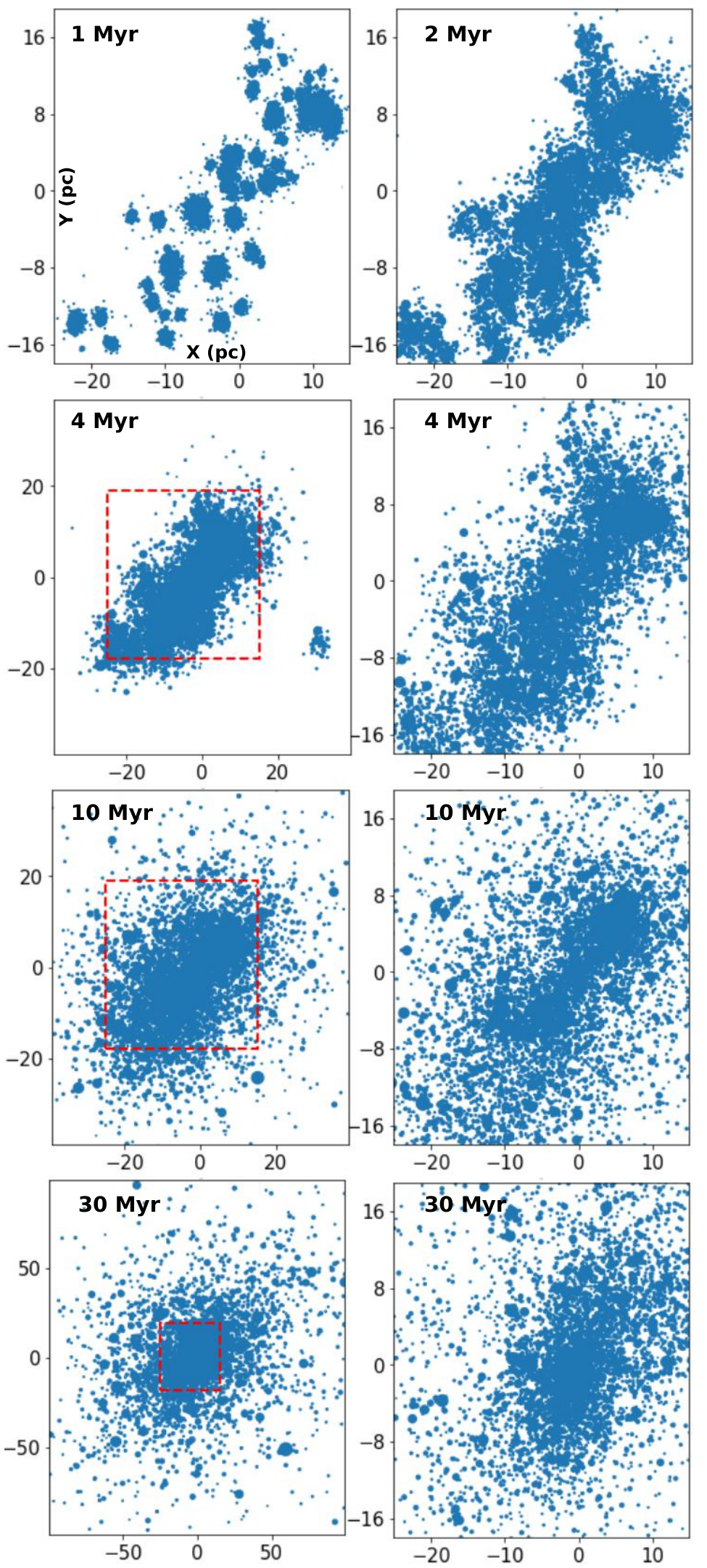}
\caption{The snapshots on the XY plane in the case of "Carina-fast" starting from Fig.~\ref{example1}(b). The size of the points in the map represents the stellar mass. The figure on the right shows the region within the red dashed box in the figure on the left.}
\label{Carina-fast}
\end{figure}
\begin{figure}
\centering
\includegraphics[width=0.45\textwidth]{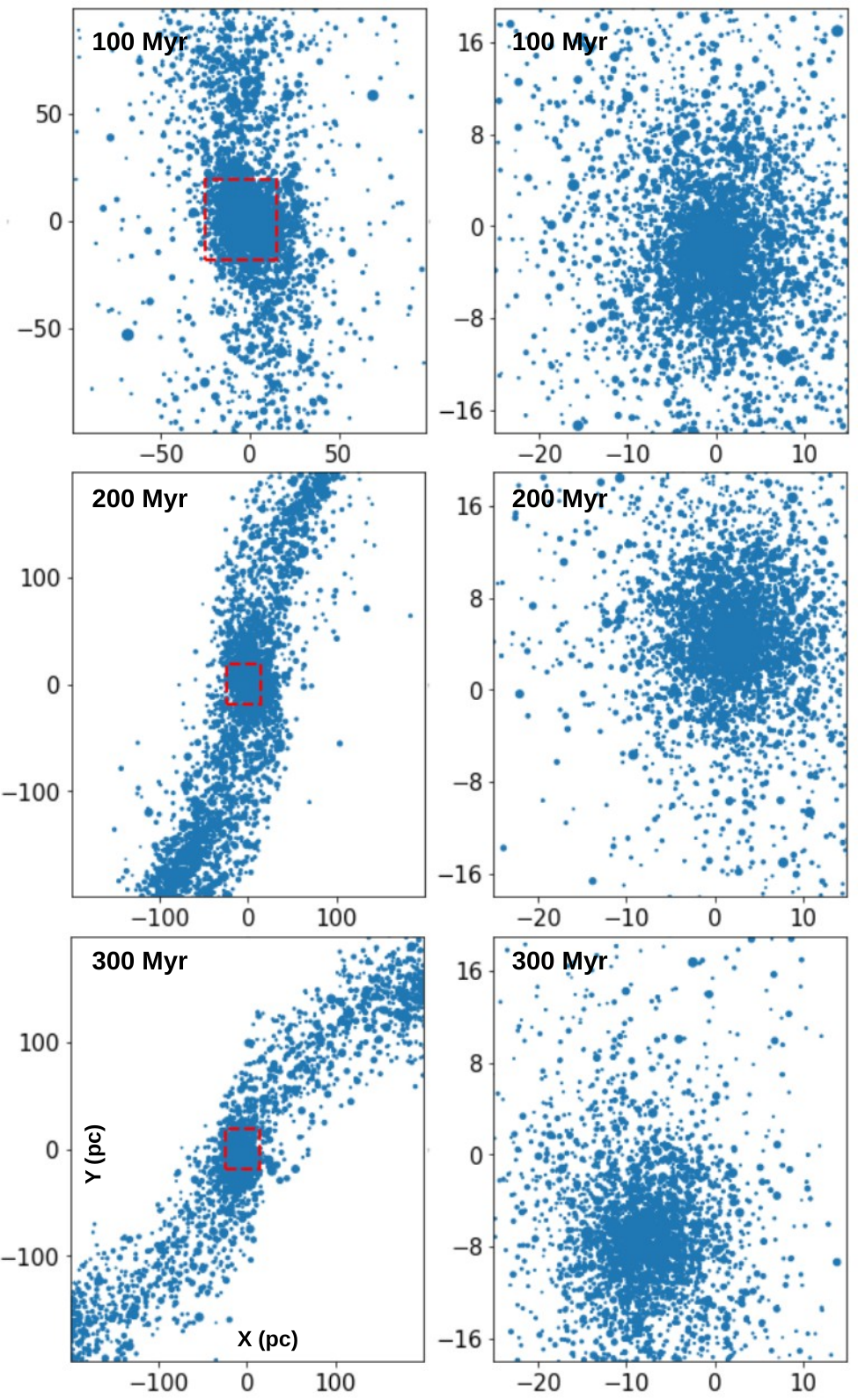}
\caption{Continued Fig.~\ref{Carina-fast}.}
\label{Carina-fast1}
\end{figure}

\begin{figure}
\centering
\includegraphics[width=0.45\textwidth]{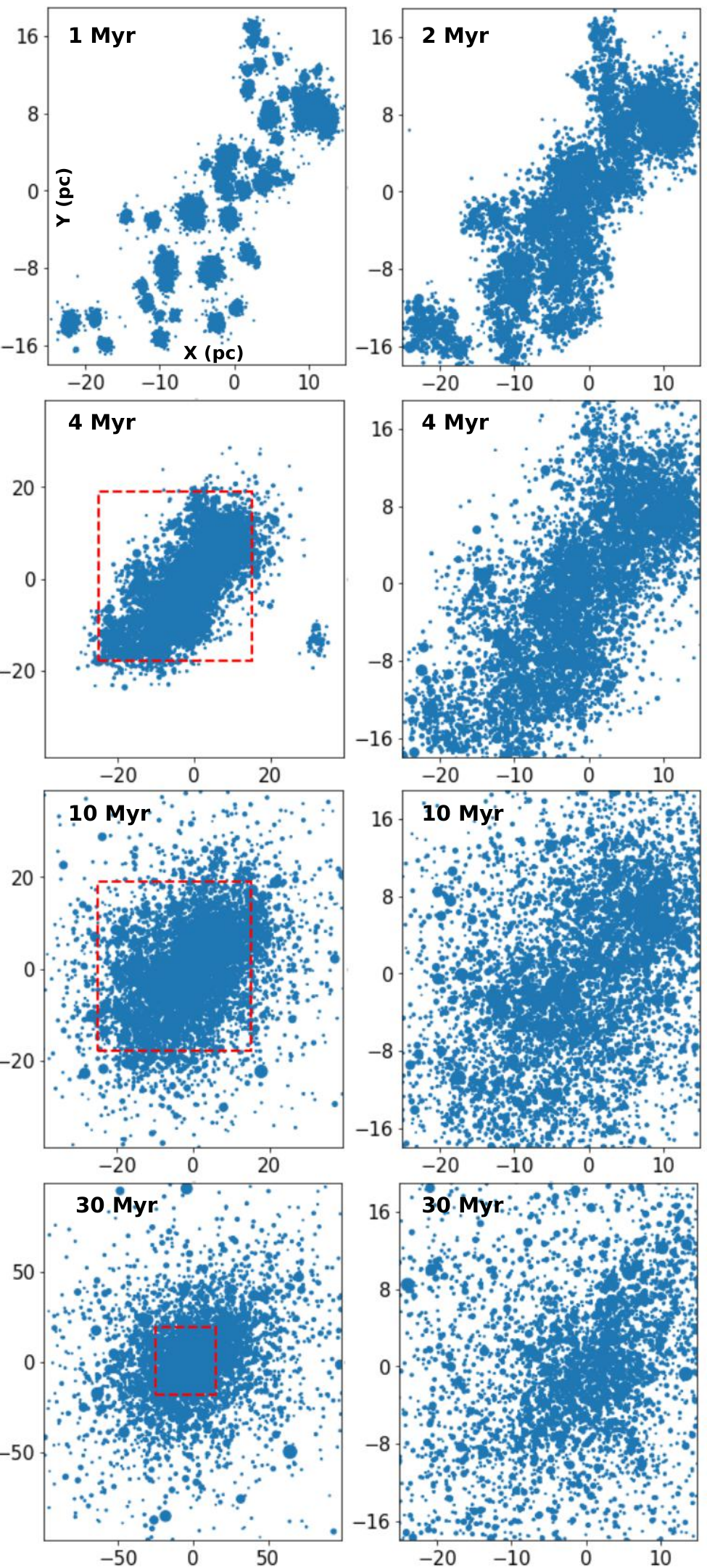}
\caption{The snapshots on the XY plane in the case of "Carina-fast-vd" starting from Fig.~\ref{example1}(c). The size of the points in the map represents the stellar mass. The figure on the right shows the region within the red dashed box in the figure on the left.}
\label{Carina-fast-vd}
\end{figure}

\begin{figure}
\centering
\includegraphics[width=0.45\textwidth]{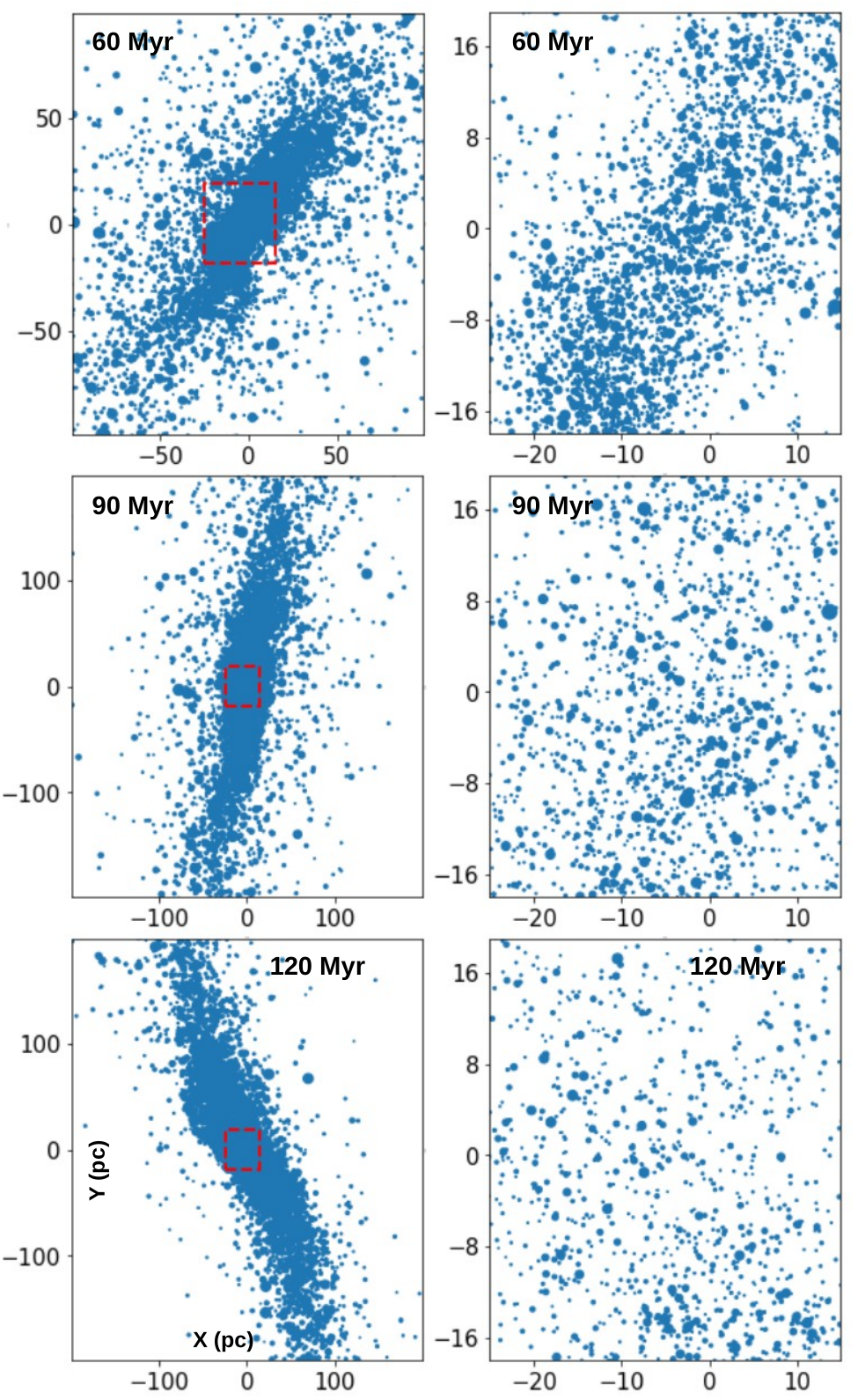}
\caption{Continued Fig.~\ref{Carina-fast-vd}.}
\label{Carina-fast-vd1}
\end{figure}

\end{appendix}

\clearpage
\noindent
\end{document}